\documentclass{emulateapj}

\shorttitle{Comment on arXiv:1006.0488}
\shortauthors{Alister W.\ Graham}
\slugcomment{submitted to arXiv.org/astro-ph}

\begin{document}

\title{Comment on arXiv:1006.0488}

\author{Alister W.\ Graham}
\affil{Centre for Astrophysics and Supercomputing, Swinburne University
of Technology, Hawthorn, Victoria 3122, Australia.}

\begin{abstract} 

This comment is in response to the article titled ``A Non-Parametric Estimate
of Mass Scoured in Galaxy Cores" (http://arxiv.org/abs/1006.0488) written by
Philip F.\ Hopkins and Lars Hernquist.  It politely mentions two relevant
papers in which the main conclusion from Hopkins \& Hernquist had already been
published six years ago using the core-S\'ersic model.  It then explains why
Hopkins \& Hernquist's concern about the core-S\'ersic model is not
appropriate.

\end{abstract}


\section{Comment}

Graham (2004) revealed, using the core-S\'ersic model
(Graham et al.\ 2003; Trujillo et al.\ 2004), that the central, stellar mass
deficit in massive elliptical galaxies is around 1-2 times the central black
hole mass (see Graham 2004, his Figure~2). Graham (2004) emphasized that this deficit
was some 10 times smaller than 
previous estimates.  Quoting Graham (2004), it reads "These results are in agreement with the
theoretical expectations of mass ejection from binary black hole mergers and
also with popular $\Lambda$CDM models of hierarchical galaxy formation" and he 
noted that "Theoretical expectations for the ejected core mass, after the violent
unification of galaxies containing [supermassive black holes] SMBHs, scale as 0.5-2$NM_{\rm bh}$, where
$M_{\rm bh}$ is the {\it final} BH mass and $N$ is the number of merger events
(Milosavljevi\'c et al.\ 2002)."

Trujillo et al.\ (2004) applied the core-S\'ersic model to massive
elliptical galaxies and reported that the sizes of the depleted cores were
several time smaller than previously reported, and, as can be seen in their 
Figure~8 (and their Tables~2 and 4), that the core radii roughly span 20 to 100 pc. 

Hopkins \& Hernquist (arXiv:1006.0488) write in the second last paragraph of
their Introduction that the central mass deficits rise ``until asymptoting to
a maximum scoured mass $\sim$1−-2 $M_{\rm BH}$ near $\sim$100 pc.''  In their
Section~3 they write that ``between $\sim$10-−100 pc, this appears to
asymptote to a maximum of $\Delta M \sim$1-−2 $M_{\rm BH}$'', and in the
fourth paragraph of their Discussion that ``The apparent scoured mass then
grows [snip] until $\sim$10--100 pc at which point it asymptotes to a constant
mass fraction $\sim$1--2 $M_{\rm BH}$''.\footnote{The abstract of Hopkins \&
Hernquist appears to accidentally state a mass range from 0.5--2 $M_{\rm BH}$, rather
than the value of 1--2 $M_{\rm BH}$ reported in their paper.}

The fifth paragraph of the Discussion in Hopkins \& Hernquist reads 
"These results are consistent with the expectation from $N$-body
experiments of the effects of a binary BH on the central stellar
mass distribution. Such experiments typically find that the stellar
mass scattered before the binary BH merges is $\sim$0.5−-1.5 $M_{\rm BH}$".
Their abstract thus reports that "The relatively low mass deficits
inferred, and characteristic radii, are in good agreement with models of
scouring from BH binary systems".

While the nice analysis by Hopkins \& Hernquist confirms that performed by Graham
(2004) and Trujillo et al.\ (2004), it does highlight that their concern about
past application of the core-S\'ersic model to elliptical galaxies is not valid.

\section{Mis-fitting the core-S\'ersic model}

Section~2 of Hopkins \& Hernquist covers a topic raised by Hopkins et al.\
(2008, 2009), illustrating why those who have used the inward extrapolation of
a S\'ersic model might have got things wrong {\it if} elliptical galaxies have
distinct inner and outer (violently relaxed) components.  While in theory this
concern is a good one, in reality/practice it is not an appropriate criticism
of past works which used the core-S\'ersic model.  The lower panel of Figure~1
in Hopkins \& Hernquist presents a ``false'' mass deficit, derived using the
core-S\'ersic model applied to a simulated, two-component, galaxy light
profile; the fit yields a core radius of 470 pc and mass deficit which is 5\%
of the total stellar mass.  However, these values for a depleted core are much
higher than the core radii in Trujillo et al.\ (2003, $\lesssim$100 pc) and
the 0.07-0.7\% mass deficits reported by Graham (2004).  That is, this is not
the type of core that has been found when using the core-S\'ersic model
applied to elliptical galaxies, and as such Section~2 of Hopkins \& Hernquist
inappropriately blankets the core-S\'ersic model as "not to be trusted".

Trujillo et al.\ (2004, their Figure~3) discussed how to read the nature of
residual profiles (model-data) and reported a typical residual scatter about
the fitted core-S\'ersic model of 0.05 mag arcsec$^{-2}$.  The core-S\'ersic
model fit by Hopkins et al.\ (2008b, their Figure~2) and the residual profile
in Hopkins et al.\ (2009, their Figure~30) clearly reveal that their fitted
core-S\'ersic model is nothing like the correctly fit core-S\'ersic models in
Trujillo et al.\ (2004).  Light profiles such as that shown in Figure~1 of
Hopkins \& Hernquist {\it are} modelled with two components by observers who
use the S\'ersic model (e.g.\ Seigar, Graham \& Jerjen 2007).  As it is,
Figure 1 from Hopkins \& Hernquist is rather misleading.  Thankfully, past
application of the core-S\'ersic model to real elliptical galaxies does not
suffer from the issue presented in this Figure, which is why Hopkins \&
Hernquist's model-independent results agree exactly with the results reported
by Graham (2004) and Trujillo et al.\ (2004).

\end{document}